\documentclass[aps,showpacs]{revtex4}
\usepackage{graphicx}
\usepackage{color}
\usepackage[T1]{fontenc}
\usepackage[normalem]{ulem}

\RequirePackage{amsfonts,amssymb,amsmath}

\begin{document}
\title{  n-Dimensional FLRW   Quantum Cosmology   }
\author{Patricio S. Letelier} 
 \email{e-mail: letelier@ime.unicamp.br} 
\author{Jo\~ao Paulo M. Pitelli} 
\email{e-mail:pitelli@ime.unicamp.br}
\affiliation{
Departamento de Matem\'atica Aplicada-IMECC,
Universidade Estadual de Campinas,
13081-970 Campinas,  Sao Paulo, Brazil}
%-----------------------------------------------------OK-----------------------------------------------------------------------------------------------

%-----------------------------------------------------------Abstract-----------------------------------------------------------------------------------
%-----------------------------------------------------OK-----------------------------------------------------------------------------------------------
\begin{abstract}
We introduce the formalism of quantum cosmology in a Friedmann-Lema\^itre-Robertson-Walker (FLRW) universe of arbitrary dimension filled with a perfect fluid with   $p=\alpha\rho$ equation of state. First we show that the Schutz formalism, developed in four dimensions, can be extended to a  n-dimensional universe.  We compute  the quantum representant of the scale factor $a(t)$,  in the Many-Worlds, as well as, in the de Broglie-Bohm interpretation of quantum mechanics. We show that the singularities, which are still present in the n-dimensional generalization of FLRW universe, are excluded with the introduction of quantum theory. We quantize, via the de Broglie-Bohm interpretation of quantum mechanics, the components of the Riemann curvature tensor in a  tetrad basis in a n-dimensional FLRW universe filled with radiation ($p=\frac{1}{n-1}\rho$).  We show that the quantized version of the Ricci scalar are perfectly regular for all time $t$. We also study the behavior of the energy density and pressure and show that the ratio $\left<p\right>_L/\left<\rho\right>_L$ tends to the classical value $1/(n-1)$ only for $n=4$, showing that $n=4$ is somewhat privileged among the other dimensions. Besides that, as $n\to\infty$, $\left<p\right>_L/\left<\rho\right>_L\to 1$. 
\end{abstract}

\pacs{04.50.-h, 04.60.-m, 98.80.Qc}
%-----------------------------------------------------OK-----------------------------------------------------------------------------------------------
\maketitle

%-----------------------------------------------------OK-----------------------------------------------------------------------------------------------
\section{Introduction}

Recently there has been a current interest in the study of models that involve extra dimensions. The first works on this subject date back to the 1920s with Kaluza and his assumption that spacetime is five-dimensional in order to unify gravity with the electromagnetism. The inclusion of  the electromagnetic potential  within a geometrical picture as components of the metric tensor along the fifth dimension, leads to the Einstein-Maxwell equations, which follows from  the vacuum Einstein equations in five dimensions. The nonobservability of the extra dimension  was 
considered  by Klein, who showed that this would explained if  the extra dimension is compatified, having a circular topology and a small enough scale (of the order of the Planck length).  This tiny radius of curvature implies that the extra dimension would remain hidden to all low energy physics considerations \cite{pérez-lorenzana}. Over the years, the Kaluza-Klein idea was generalized to higher dimensions in order to unify non-Abelian gauge fields with gravity \cite{cho1,cho2} and now it is believed that spacetime could have much more than four dimensions, such as in the 10-dimensional superstring \cite{green} theory and in the 11-dimensional supergravity \cite{buchbinder}. 

One of the main problems with the 4-dimensional Friedmann-Lema\^itre-Robertson-Walker (FLRW) universe filled with a perfect fluid is the presence of the big-bang singularity. Such a singularity is a scalar curvature singularity, where every observer near the singularity sees physical quantities diverging \cite{helliwell2, konkowski}. It was shown \cite{hawking} that under very reasonable assumption, singularities are always present in cosmology. Near the singularity, the laws of physics break-down and we hope that a full quantum theory of gravity will overcome this situation. While such a theory does not exist, there are many attempts to incorporate quantum mechanics into general relativity. One of the first attempts to do this was quantum cosmology \cite{dewitt}. The introduction of quantum cosmology in the minisuperspace of FLRW universe has been shown successfully in the sense that in the Many-Worlds and in the de Broglie-Bohm interpretation of quantum mechanics, the singularity are  removed \cite{alvarenga, lemos, oliveira neto}. 

In the n-dimensional FLRW universe, the big-bang singularity is still present. So in this paper we follow the idea that our physical four-dimensional Universe is actually embedded in a higher-dimensional spacetime and generalize the concept of quantum cosmology to a flat, homogeneous, and isotropic n-dimensional FLRW universe in order to see if the singularities are excluded with the introduction of quantum mechanics. The generalization of classical inflationary cosmological solutions for a n-dimensional FLRW universe has already been made \cite{garcia} and it was shown that the derived solutions possess properties similar to those ones of the corresponding solutions of the standard four-dimensional FLRW cosmology. Here we show that, like in the 4-dimensional FLRW quantum cosmology, the singularity in the n-dimensional universe is removed with the introduction of quantum cosmology and that the expectation value of the scale factor in the Many-Worlds interpretation, and the trajectory of the scale factor in the de Broglie-Bohm interpretation, tends to the classical behavior as $t\to\infty$. All of our results reduces to the well known results for $n=4$. We also study the local expectation value of the energy density and pressure of the fluid for a universe filled with radiation and show that the ratio $\left<p\right>_L/\left<\rho\right>_L$ tends to the classical limit as $t\to\infty$ only for the $n=4$ case, fact that single out four dimensions.

In this paper we proceed as follows: in Sec. II   we generalize Schutz formalism \cite{schutz1, schutz2} to n-dimensional fluids  and solve the  Einstein equations  via Hamiltonian formalism to   find the scale factor for the n-dimensional FLRW universe filled with a perfect fluid with equation of state $p=\alpha\rho$.  We compare this scale factor with  the one found via Einstein equations. In Sec. III we solve the Wheeler-deWitt equation of the universe,  find the expectation value and the trajectory of the scale factor and compare with the classical value. In Sec. IV we quantize the independent components of the Riemann curvature tensor in a tetrad basis in a radiation-dominated universe and show that the Ricci scalar is perfectly regular for all time $t$, illustrating the exclusion of the classical singularity with the introduction of quantum theory. We also quantize the energy density and pressure of the fluid and study the behavior of the ratio $p/\rho$ with relation to time. Finally, in Sec. V, we discuss the main results presented in this work.

\section{Classical formalism}
\subsection{Einstein equations}
Let us find the scale factor in the n-dimensional flat FLRW universe filled with a perfect fluid through the Einstein equations. A flat n-dimensional FLRW universe has the metric
\begin{equation}
ds^2=-dt^2+a^2(t)\delta_{ij}dx^{i}dx^{j}\;\;\;\;\;(i,j=1\dots n-1),
\end{equation}
where $\delta_{ij}$ is the Kronecker delta symbol and $x^i$ are the comoving coordinates of the universe. In this coordinate set, the n-velocity of the fluid which fills the universe is given by
\begin{equation}
U^{\mu}=\left(1,0,\dots,0\right),
\end{equation} 
so that the energy momentum tensor 
\begin{equation}
T_{\mu\nu}=(\rho+p)U_{\mu}U_{\nu}+pg_{\mu\nu}
\end{equation}
takes the form
\begin{equation}
T_{\mu\nu}=\text{diag}(\rho, a^2p, \dots, a^2p).
\end{equation}

The Einstein equations for an n-dimensional spacetime are given by \cite{garcia}
\begin{equation}
G_{\alpha\beta}=R_{\alpha\beta}-\frac{1}{2}g_{\alpha\beta}R=\kappa_n T_{\alpha\beta},
\end{equation}
with $\kappa_n$ representing the multidimensional gravitational constant. The independent Einstein equations are 
\begin{subequations}
\begin{align}
&G_{00}=\kappa_n T_{00}\Rightarrow \frac{(n-1)(n-2)}{2}\left(\frac{\dot{a}}{a}\right)^2=\kappa_n \rho,\label{eqa}\\
&G_{ii}=\kappa_n T_{ii}\Rightarrow -\frac{(n-2)(n-3)}{2}\dot{a}^2-(n-2)\ddot{a}a=\kappa_n a^2 p\;\;\;\;\;(i=1,\dots,n-1)\label{eqb}.
\end{align}
\end{subequations}

These are the Friedmann equations for a flat n-dimensional FLRW universe filled with a perfect fluid. Note that there are only two independent equation due to isotropy and homogeneity. By introducing an equation of state for the fluid $p=\alpha\rho$, we can replace Eq. (\ref{eqa}) into Eq. (\ref{eqb}) obtaining
\begin{equation}
\frac{1}{2}\left[\alpha (n-1)+(n-3)\right]\frac{\dot{a}}{a}+\frac{\ddot{a}}{\dot{a}}=0.
\end{equation}

The above equation can be integrated, 
\begin{equation}
\ln{a^{\frac{\alpha (n-1)+(n-3)}{2}}\dot{a}}=\ln{\text{const}}.
\end{equation}
Therefore
\begin{equation}
\dot{a}\propto a^{-\frac{\alpha (n-1)+(n-3)}{2}}\Rightarrow a\propto t^{\frac{2}{(n-1)(1+\alpha)}}.
\label{scale factor}
\end{equation}

This is the scale factor for a flat n-dimensional FLRW universe filled with a perfect fluid with equation of state $p=\alpha \rho$. The big-bang singularity is present no matter what dimension we take.  Note that for $n=4$ and $\alpha=1/3$ (a four-dimensional flat radiation-dominated universe) we have $a\propto t^{1/2}$, and for $\alpha=0$ (matter-dominated universe) we have $a\propto t^{2/3}$ as required.

\subsection{Hamiltonian formalism}

Now we will develop the Hamiltonian formalism for the n-dimensional FLRW universe filled with a perfect fluid with equation of state $p=\alpha\rho$. We will generalize Schutz formalism \cite{schutz1,schutz2} to n-dimensions and check its validity by  calculating the scale factor via a variational principle and comparing with the scale factor calculated via Friedmann equations in the previous subsection.

The action for general relativity is given by
\begin{equation}
S_{GR}=\int_{\mathcal{M}}{d^{n}x\sqrt{-g}R}+2\int_{\partial \mathcal{M}}{d^{n-1}x\sqrt{h}h_{ab}K^{ab}},
\label{action for general relativity}
\end{equation}
where $h_{ab}$ is the induced metric over the boundary $\partial \mathcal{M}$ of the four-dimensional manifold $\mathcal{M}$ and $K^{ab}$ is the extrinsic curvature of the hypersurface $\partial \mathcal{M}$, that is, the curvature of $\partial \mathcal{M}$ with respect to $\mathcal{M}$.

In a  four-dimensional spacetime, Schutz showed that the four-velocity of a perfect fluid is expressed in terms of five potentials \cite{schutz1}
\begin{equation}
U_{\nu}=\mu^{-1}(\phi_{,\nu}+\alpha\beta_{,\nu}+\theta S_{,\nu}),
\label{four-velocity}
\end{equation}
where $\mu$ is the specific enthalpy and $S$ is the specific entropy. The potentials $\alpha$ and $\beta$ are connected with rotation \cite{schutz1} so they are not present in the FLRW universe due to its symmetry. The potentials $\phi$ and $\theta$ have no clear physical meaning and the four-velocity satisfies the normalization condition
\begin{equation}
U_{\nu}U^{\nu}=-1.
\end{equation}

Schutz also showed that the action for the fluid is given by
\begin{equation}
S_{f}=\int_{\mathcal{M}}{d^4x\sqrt{-g}p},
\end{equation}
where $p$ is the pressure of the fluid and is linked with the density by the equation of state $p=\alpha\rho$.

By thermodynamical considerations, Lapchinskii and Rubakov \cite{lapchinskii} found that the expression for the pressure is given in terms of the potentials by
\begin{equation}
p=\frac{\alpha}{(\alpha+1)^{1/\alpha+1}}(\dot{\phi}+\theta\dot{S})^{1/\alpha+1}\exp{\left(-\frac{S}{\alpha}\right)},
\end{equation}
so that the action for the four-dimensional flat FLRW universe with metric
\begin{equation}
ds^2=-N(t)^2dt^2+a^2(t)(dx^2+dy^2+dz^2),
\end{equation}
where $N(t)$ is the lapse function, is given by \cite{alvarenga} (in units where $16\pi G=1$)
\begin{equation}
S=\int{dt\left[-6\frac{\dot{a}^2a}{N}+N^{-1/\alpha}a^3\frac{\alpha}{(\alpha+1)^{1/\alpha+1}}(\dot{\phi}+\theta\dot{S})^{1/\alpha+1}\exp{\left(-\frac{S}{\alpha}\right)}\right]}.
\end{equation}

In an n-dimensional spacetime, more potentials should be added in Eq. (\ref{four-velocity}). We suppose that these new potentials, just like $\alpha$ and $\beta$ in Eq. (\ref{four-velocity}), are related with generalized rotations and are absent of the n-dimensional FLRW universe. Therefore the action for the flat n-dimensional FLRW universe filled with a perfect fluid with equation of state $p=\alpha\rho$ is given by
\begin{equation}
S=\int{dt L}=\int{dt\left[-(n-1)(n-2)\frac{\dot{a}^2a^{n-3}}{N}+N^{-1/\alpha}a^{n-1}\frac{\alpha}{(\alpha+1)^{1/\alpha+1}}(\dot{\phi}+\theta\dot{S})^{1/\alpha+1}\exp{\left(-\frac{S}{\alpha}\right)}\right]}.
\end{equation}

The momenta associated with $a$, $\phi$ and $S$ are given by
\begin{subequations}
\begin{align}
&p_a=\frac{\partial L}{\partial \dot{a}}=-2(n-1)(n-2)\frac{\dot{a}a^{n-3}}{N}\label{eqa1},\\
&p_{\phi}= \frac{\partial L}{\partial \dot{\phi}}=N^{-1/\alpha}a^{n-1}\frac{(\dot{\phi}+\theta\dot{S})^{1/\alpha}}{(1+\alpha)^{1/\alpha}}e^{-S/\alpha}\label{eqb1},\\
&p_{S}=\frac{\partial L}{\partial \dot{S}}=\theta p_\phi.
\end{align}
\end{subequations}
The Hamiltonian of the system is then given by
\begin{equation}
H=p_a \dot{a}+p_\phi(\dot{\phi}+\theta \dot{S})-L=N\left[-\frac{p_a^2}{4(n-1)(n-2)a^{n-3}}+a^{(n-1)\alpha}e^{S}p_{\phi}^{\alpha+1}\right].
\end{equation}

After a canonical transformation of the form \cite{lapchinskii,alvarenga}
\begin{equation}
T=-p_Se^{-S}p_\phi^{-(\alpha+1)},\;\;\;p_T=p_\phi^{\alpha+1},\;\;\; \overline{\phi}=\phi-(\alpha+1)\frac{p_S}{p_\phi},\;\;\; \overline{p}_\phi=p_\phi,
\end{equation}
and, since the lapse function $N$ plays the role of a Lagrange multiplier in the formalism, we have that the super-Hamiltonian of the system reads
\begin{equation}
\mathcal{H}=-\frac{p_a^2}{4(n-1)(n-2)a^{n-3}}+\frac{p_T}{a^{(n-1)\alpha}}\approx 0,
\end{equation}
where $p_T$ is a new canonical variable conjugated to dynamical degree of the fluid. Having the Hamiltonian of the system we can extract the equations of motion. They are \cite{barbozajr.}
\begin{equation}
\begin{aligned}
&\dot{a}=\frac{\partial(N\mathcal{H})}{\partial p_a}=-\frac{Np_a}{2(n-1)(n-2)a^{n-3}},\;\;\; \dot{p}_a=-\frac{\partial(N\mathcal{H})}{\partial a}=-\frac{N(n-3)}{4(n-1)(n-2)}\frac{p_a^2}{a^{n-2}}+\frac{\alpha(n-1)Np_T}{a^{(n-1)\alpha}+1},\\
&\dot{T}=\frac{\partial(N\mathcal{H})}{\partial p_T}=\frac{N}{a^{(n-1)\alpha}},\;\;\;\dot{p}_T
=-\frac{\partial(N\mathcal{H})}{\partial T}=0.\end{aligned}
\label{equations of motion}
\end{equation}

By the super-Hamiltonian constraint we have 
\begin{equation}
p_a=-2\sqrt{(n-1)(n-2)p_T}a^{-[(n-1)\alpha-(n-3)]/2},
\end{equation}
where we chose the negative sign in order to have an expanding universe \cite{barbozajr.}. Therefore, in the cosmic gauge $N=1$ the equation for $a(t)$ reads
\begin{equation}
\dot{a}=a^{-[(n-1)\alpha+(n-3)]/2}\sqrt{\frac{p_T}{(n-1)(n-2)}},
\end{equation}
whose solution is
\begin{equation}
a^{(n-1)(1+\alpha)/2}=a_0^{(n-1)(1+\alpha)/2}+\sqrt{\frac{p_T}{4(n-1)(n-2)}}(n-1)(1+\alpha)(t-t_0),
\end{equation}
where $a_0$ is the present value of the scale factor. Hence
\begin{equation}
a(t)=\left[a_0^{(n-1)(1+\alpha)/2}+\sqrt{\frac{p_T}{4(n-1)(n-2)}}(n-1)(1+\alpha)(t-t_0)\right]^{\frac{2}{(n-1)(1+\alpha)}},
\end{equation}
and by a suitable redefinition of time we have 
\begin{equation}
a(t)\propto t^{\frac{2}{(n-1)(1+\alpha)}}
\end{equation}
just like Eq. (\ref{scale factor}), fact that validates our assumptions.

\begin{figure}[!htb]
\centering
\includegraphics[scale=0.8]{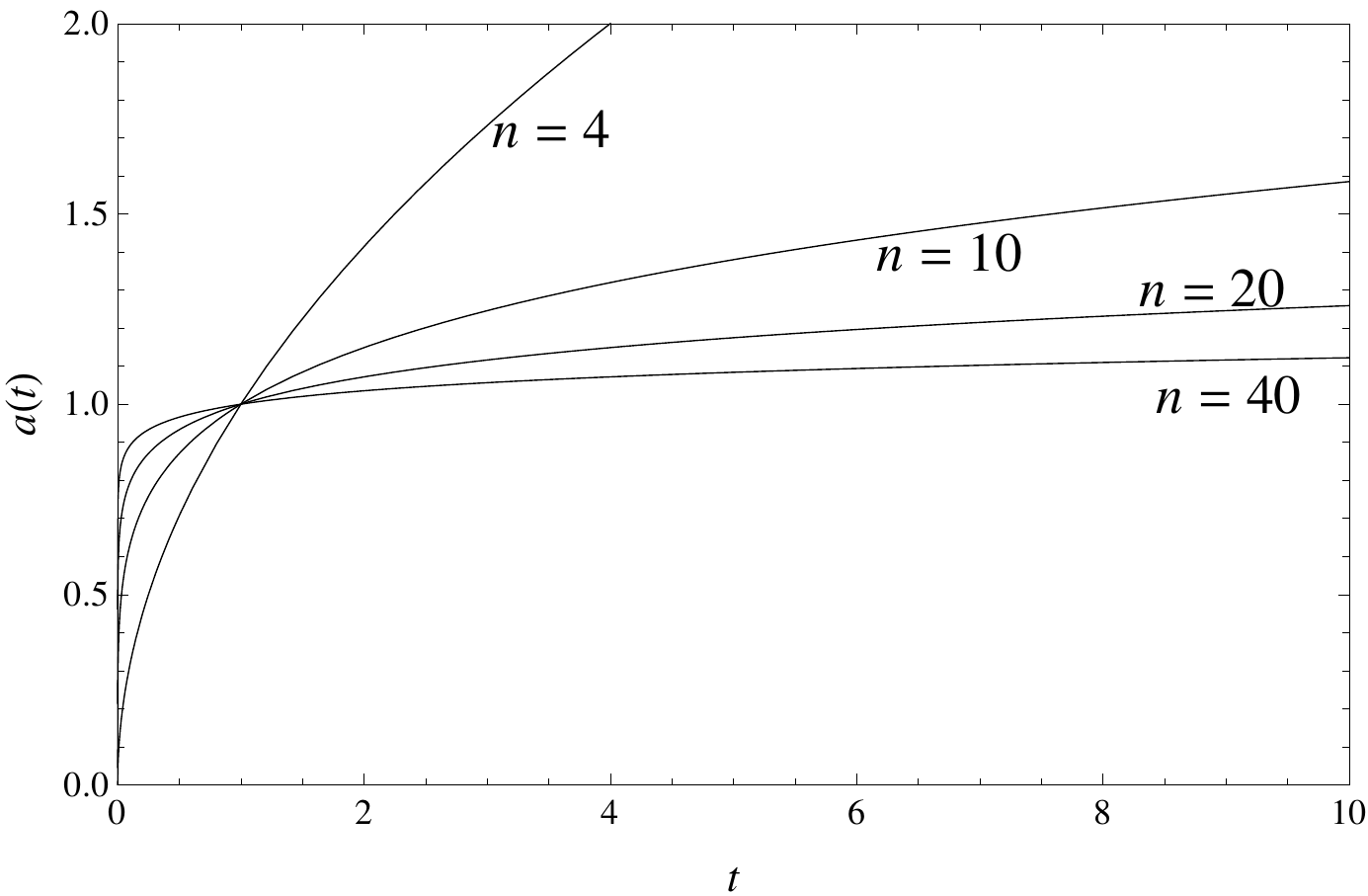}
\caption{The scale factor for $n=4,10,20$ and $40$ (for $\alpha=1/(n-1)$). We note that the big-bang is still present.}
\label{a(t) classic}
\end{figure}

In Fig. \ref{a(t) classic} we show the behavior of the scale factor for $n=4,10,20$ and $40$ in the radiation-dominated era $\alpha=1/(n-1)$. We see that as $n\to\infty$ the universe tends to a static universe.

\section{Wheeler-deWitt equation for the flat n-dimensional FLRW universe with a perfect fluid}

By following the Dirac approach \cite{dirac} for quantization of Hamiltonian systems with constraints, making the substitutions
\begin{equation}
p_a\to-i\frac{\partial}{\partial a}, \;\;\; p_T=-i\frac{\partial}{\partial T}
\end{equation}
and demanding
\begin{equation}
\mathcal{H}\Psi=0,
\end{equation}
where $\Psi$ is the wave function of the universe, we have
\begin{equation}
\frac{a^{(n-1)\alpha-(n-3)}}{4(n-1)(n-2)}\frac{\partial^2\Psi}{\partial a^2}+i\frac{\partial \Psi}{\partial t}=0,
\label{eq. universe}
\end{equation}
with $t=-T$ being the time coordinate in the conformal gauge $N=a^{(n-1)\alpha}$ [see Eq. (\ref{equations of motion})].

This equation has the form of a Schr\"odinger equation $i\frac{\partial\Psi}{\partial t}=\hat{H}\Psi$ with
\begin{equation}
\hat{H}=-\frac{a^{(n-1)\alpha-(n-3)}}{4(n-1)(n-2)}\frac{\partial^2}{\partial a^2}. 
\end{equation}
In order to ensure self-adjointness of the operator $\hat{H}$ above, the inner product between two state equations is defined by \cite{lemos}
\begin{equation}
\left<\Phi, \Psi\right>=\int_{0}^{\infty}{a^{(n-3)-(n-1)\alpha}\Phi^{\ast}\Psi da}.
\end{equation}
Moreover, boundary conditions near $a=0$ on the wave function must be imposed. The most common ones are 
\begin{subequations}\begin{align}
\Psi(0,t)&=0\;\;\;\;\; (\text{Dirichlet boundary condition})\label{eqa2},\\
\left.\frac{\partial\Psi}{\partial a}\right|_{a=0}&=0 \;\;\;\;\; (\text{Neumann boundary condition})\label{eqb2}.
\end{align}\end{subequations}
In the course of the paper we will choose for simplicity the Dirichlet boundary condition. Let us now solve Eq. (\ref{eq. universe}). By a separation of variables of the form $\Psi(a,t)=e^{-iEt}\psi(a)$ we have
\begin{equation}
\psi''(a)+4E(n-1)(n-2)a^{[(n-3)-(n-1)\alpha]}\psi(a)=0.
\label{bessel eq.}
\end{equation}
The solution of the above equation, respecting the Dirichlet boundary condition, is given by \cite{bessel}
\begin{equation}
\psi_E(a)=a^{1/2}J_{\frac{1}{(n-1)(1-\alpha)}}\left(\frac{4\sqrt{E}}{(1-\alpha)}\sqrt{\frac{n-2}{n-1}}a^{\frac{(n-1)(1-\alpha)}{2}}\right),
\end{equation}
where $J_{\nu}(x)$ is the Bessel function of order $\nu$. Note that this solution is valid only for $\alpha\neq 1$. For $\alpha=1$, solution of Eq. (\ref{bessel eq.}) is given by $a^{\frac{1\pm\sqrt{1-16E(n-1)(n-2)}}{2}}$, which diverge either at $a=0$ or $a=\infty$ (note that $E$ is a variable to be integrated in order to find wave packets representing the wave function of the universe).

From these stationary states we can find finite norm wave packets by superposing states of the form \cite{alvarenga}
\begin{equation}
\Psi(a,t)=a^{1/2}\int_{0}^{\infty}{dE A(E)e^{-iEt}J_{\frac{1}{(n-1)(1-\alpha)}}\left(\frac{4\sqrt{E}}{(1-\alpha)}\sqrt{\frac{n-2}{n-1}}a^{\frac{(n-1)(1-\alpha)}{2}}\right)}.
\end{equation}

Defining $u=\sqrt{E}$ we have
\begin{equation}
\Psi(a,t)=2a^{1/2}\int_0^\infty{du A(u)ue^{-itu^2}J_{\frac{1}{(n-1)(1-\alpha)}}\left(\frac{4u}{(1-\alpha)}\sqrt{\frac{n-2}{n-1}}a^{\frac{(n-1)(1-\alpha)}{2}}\right)}.
\end{equation}

Let us take a quasi-Gaussian superposition factor $A(u)=u^{\frac{1}{(n-1)(1-\alpha)}}e^{-\gamma u^2}/2$. We  have

\begin{equation}
\Psi(a,t)=2a^{1/2}\int_0^\infty{du u^{\frac{1}{(n-1)(1-\alpha)}+1}e^{-(\gamma+ it)u^2}J_{\frac{1}{(n-1)(1-\alpha)}}\left(\frac{4u}{(1-\alpha)}\sqrt{\frac{n-2}{n-1}}a^{\frac{(n-1)(1-\alpha)}{2}}\right)}.
\end{equation}

Now we use the following  relation \cite{gradshteyn}
\begin{equation}
\int_0^\infty{x^{\nu+1}e^{-\alpha u^2}J_{\nu}(\beta x)dx}=\frac{\beta^{\nu}}{(2\alpha)^{\nu+1}}\exp{\left(-\frac{\beta^2}{4\alpha}\right)}
\end{equation}
to obtain
\begin{equation}\begin{aligned}
\Psi(a,t)&=\frac{\left[\frac{4}{(1-\alpha)}\sqrt{\frac{n-2}{n-1}}\right]^{\frac{1}{(n-1)(1-\alpha)}}}{[2(\gamma+it)]^{\frac{1+(n-1)(1-\alpha)}{(n-1)(1-\alpha)}}}a\exp{\left[-\frac{4}{(1-\alpha)^2(\gamma+it)}\left(\frac{n-2}{n-1}\right)a^{(n-1)(1-\alpha)}\right]}\\
&=\frac{\text{const.}}{[2(\gamma+it)]^{\frac{(n-1)(1-\alpha)+1}{(n-1)(1-\alpha)}}}a\exp{\left[-\frac{4\gamma}{(1-\alpha)^2}\left(\frac{n-2}{n-1}\right)\frac{a^{(n-1)(1-\alpha)}}{\gamma^2+t^2}\right]}\exp{\left[\frac{4it}{(1-\alpha)^2}\left(\frac{n-2}{n-1}\right)\frac{a^{(n-1)(1-\alpha)}}{\gamma^2+t^2}\right]}.
\end{aligned}\end{equation}

By adopting the many-worlds interpretation of quantum mechanics, we can calculate the expectation value of the scale factor
\begin{equation}
\left<a\right>(t)=\frac{\left<\Psi\left|a\right|\Psi\right>}{\left<\Psi|\Psi\right>}=\frac{\int_0^\infty{a^{(n-3)-(n-1)\alpha}\Psi^{\ast}(a,t)a\Psi(a,t)da}}{\int_0^\infty{a^{(n-3)-(n-1)\alpha}\Psi^{\ast}(a,t)\Psi(a,t)da}}.
\end{equation}

Therefore 
\begin{equation}
\left<a\right>(t)\propto\frac{1}{(\gamma^2+t^2)^{\frac{1+(n-1)(1-\alpha)}{(n-1)(1-\alpha)}}}\int_0^\infty{a^{(n-3)-(n-1)\alpha}}a^3\exp{\left[-\frac{8\gamma}{(1-\alpha)^2}\left(\frac{n-2}{n-1}\right)\frac{a^{(n-1)(1-\alpha)}}{\gamma^2+t^2}\right]}.
\end{equation}
The  identity \cite{gradshteyn}
\begin{equation}
\int_0^\infty{x^{\nu-1}e^{-\mu x^p}}=\frac{1}{p}\mu^{-\frac{\nu}{p}}\Gamma\left(\frac{\nu}{p}\right)
\end{equation}
gives us
\begin{equation}
\left<a\right>(t)\propto\frac{1}{(\gamma^2+t^2)^{\frac{1+(n-1)(1-\alpha)}{(n-1)(1-\alpha)}}}\frac{1}{(\gamma^2+t^2)^{\frac{-(n+1)+(n-1)\alpha}{(n-1)(1-\alpha)}}}=(\gamma^2+t^2)^{\frac{1}{(n-1)(1-\alpha)}}.
\end{equation}

Note that $\left<a\right>(t)\neq 0$ for all $t$ so that the big-bang singularity is eliminated  with the introduction of quantum cosmology. Note also that $\left<a\right>(t)$ goes asymptotically as $\left<a\right>(t)\propto t^{\frac{2}{(n-1)(1-\alpha)}}$. Let us see that in the cosmic time  $\left<a\right>(t)$ mimics Eq. (\ref{scale factor}) in the classical limit $t\to \infty$. To do  this we remember that we are working in the conformal gauge $N=a^{(n-1)\alpha}$. Therefore,
\begin{equation}
N(t)dt=d\tau,
\end{equation}
where $\tau$ is the cosmic time. Hence
\begin{equation}
a^{(n-1)\alpha}dt=d\tau\Rightarrow t^{\frac{2(n-1)\alpha}{(n-1)(1-\alpha)}}dt=d\tau\Rightarrow t^{\frac{1+\alpha}{1-\alpha}}\propto \tau\Rightarrow t\propto \tau^{\frac{1-\alpha}{1+\alpha}}
\end{equation}
and
\begin{equation}
\left<a\right>(t)\to t^{\frac{2}{(n-1)(1-\alpha)}}\propto\tau^{\frac{2}{(n-1)(1+\alpha)}}.
\end{equation}

We see that in the classical limit, where quantum effects become negligible, the quantum scale factor reproduces the classical one as expected.

We can also calculate the scale factor through the de Broglie-Bohm interpretation  of quantum mechanics. In order to use the de Broglie-Bohm interpretation, we need to write the wave function of the universe in its polar form
\begin{equation}
\Psi=\Theta e^{iS}.
\end{equation}

In our case
\begin{equation}
\begin{aligned}
&\Theta=g(t)a\exp{\left[-\frac{4\gamma}{(1-\alpha)^2}\left(\frac{n-2}{n-1}\right)\frac{a^{(n-1)(1-\alpha)}}{\gamma^2+t^2}\right]},\\
&S=\frac{4t}{(1-\alpha)^2}\left(\frac{n-2}{n-1}\right)\frac{a^{(n-1)(1-\alpha)}}{\gamma^2+t^2}+f(t),
\end{aligned}
\label{polar form}
\end{equation}
where $g(t)$ and $f(t)$ are complicate functions that will not be necessary in what follows. Now we can calculate the Bohmian trajectories for the scale factor $a(t)$ by the use of equation \cite{bohm, holland}
\begin{equation}
p_a=\frac{\partial S}{\partial a}.
\end{equation}
Therefore, 
\begin{equation}
2(n-1)(n-2)\dot{a}\frac{a^{n-3}}{a^{(n-1)\alpha}}=\frac{4t}{(1-\alpha)^2}\left(\frac{n-2}{n-1}\right)(n-1)(1-\alpha)\frac{a^{(n-1)(1-\alpha)-1}}{\gamma^2+t^2}\Rightarrow\dot{a}=\frac{2t}{(n-1)(1-\alpha)}\frac{a}{\gamma^2+t^2},
\end{equation}
and the solution of the above equation is 
\begin{equation}
a(t)=a_0(\gamma^2+t^2)^{\frac{1}{(n-1)(1-\alpha)}}, 
\label{a(t)}
\end{equation}
where $a_0$ is a constant of integration. We see that both visions of quantum mechanics give us  to the same behavior of the scale factor.

\section{Examples}

In this section we will consider the n-dimensional FLRW universe in the radiation-dominated era. By quantizing the components of the curvature tensor in the tetrad basis via the de Broglie-Bohm interpretation of quantum mechanics, we will show that the quantized Ricci scalar is perfect regular for all times $t$, illustrating the exclusion of the big-bang singularity with the introduction of quantum cosmology. We will also find the expectation value of the energy density and pressure and study the behavior of $\left<p\right>/\left<\rho\right>$. Like we did in a previous paper \cite{pitelli} we will show that in four-dimensions $\left<p\right>/\left<\rho\right>\to 1/3$ in the classical limit. Unexpectedly, in higher dimensions this is not the case, i.e., the ratio $\left<p\right>/\left<\rho\right>$ does not tend to the classical value, indicating  a distinguish behavior for $n=4$.

Let us first consider the energy momentum tensor of a perfect fluid in n-dimensions
\begin{equation}
T_{\mu\nu}=(\rho+p)U_{\mu}U_{\nu}+pg_{\mu\nu}.
\end{equation}

In order to represent a radiation fluid we must have $T=T^{\mu}_{\phantom{\mu}\mu}=0$. Therefore
\begin{equation}
T=T^\mu_{\phantom{\mu}\mu}=-\rho+(n-1)p=0.
\end{equation}
Thus, $p=\frac{1}{n-1}\rho$ for radiation.

The non-null components of the Riemman tensor in the natural tetrad basis in n-dimensions are given by
\begin{subequations}\begin{align}
&R^{\hat{i}}_{\phantom{\hat{i}}\hat{0}\hat{i}\hat{0}}=-\frac{\ddot{a}}{a^3}+\frac{\dot{a}^2}{a^4}\;\;\;\;\; (i=1,\dots,n-1)\label{eqa3},\\
&R^{\hat{i}}_{\phantom{\hat{i}}\hat{j}\hat{i}\hat{j}}=\frac{\dot{a}^2}{a^4}\;\;\;\;\;(i,j=1,\dots,n-1;i\neq j)\label{eqb3}.
\end{align}\end{subequations}

The momentum associated with $a$ in n-dimensions in the conformal gauge $N=a$ is given by [see Eq. (\ref{eqa1})]
\begin{equation}
p_a=2(n-1)(n-2)\dot{a}a^{n-4}.
\end{equation}
Therefore the components of the Riemann tensor can be related to $a$ and $p_a$ by
\begin{subequations}
\begin{align}
&R^{\hat{i}}_{\phantom{\hat{i}}\hat{0}\hat{i}\hat{0}}=-\frac{\dot{p}_a}{2(n-1)(n-2)a^{n-1}}+(n-3)\frac{p_a^2}{4(n-1)^2(n-2)^2a^{2n-4}}\label{eqa4} \;\;\;\;\; (i=1,\dots, n-1),\\
&R^{\hat{i}}_{\phantom{\hat{i}}\hat{j}\hat{i}\hat{j}}=\frac{p_a^2}{4(n-1)^2(n-2)^2a^{2n-4}}\label{eqb4}\;\;\;\;\;(i,j=1,\dots, n-1;i\neq j).
\end{align}
\end{subequations}

Now we promote the components of the curvature tensor to the condition of quantum operators. Then we can take
the local expectation value of each component through the relation \cite{holland, oliveira neto, pitelli}
\begin{equation}
\left<R^{\hat{a}}_{\phantom{\hat{a}}\hat{b}\hat{c}\hat{d}}\right>_L=\text{Re}\left(\frac{\Psi^{\ast}\hat{R}^{\hat{a}}_{\phantom{\hat{a}}\hat{b}\hat{c}\hat{d}}\Psi}{\Psi^{\ast}\Psi}\right)
\end{equation}

Since $a$ and $p_a$ do not commute, we must choose an ordering factor in order to calculate the local expectation values of Eqs. (\ref{eqa4}) and (\ref{eqb4}). We choose the Weyl ordering \cite{lee} which is a kind of symmetrization procedure that takes all possible orders of the $a's$ and the $p's$ and then divides the result by the number of terms in the final expression. In our case we have $p_a^2/a^{2n-4}$. For the general case $f(a)p_a^2$, Weyl ordering gives us
\begin{equation}
(f(a)\hat{p}_a^2)_W=\frac{1}{4}\left[f(a)\hat{p}_a^2+2\hat{p}_af(a)\hat{p}_a+\hat{p}_a^2f(a)\right].
\label{weyl}
\end{equation}

With the aid of the following commutators, 
\begin{equation}
\begin{aligned}
&\left[a^{m},\hat{p}_a\right]=ima^{m-1}\\
&\left[a^{m},\hat{p}_a^2\right]=2ima^{m-1}\hat{p}_{a}+m(m-1)a^{m-2},
\end{aligned}
\end{equation}
we can put powers of $p_a$ at the right of powers of $a$ in Eq. (\ref{weyl}), obtaining 
\begin{equation}
\left(\frac{\hat{p}_a^2}{a^{2n-4}}\right)_W=a^{2n-4}\hat{p}_a^2+i(2n-4)a^{2n-3}\hat{p}_a-\frac{(2n-4)(2n-3)}{4a^{2n-2}}.
\end{equation}

Therefore the local expectation value of the operator $\hat{p}_a^2/a^{2n-4}$ is given by
\begin{equation}
\left<\frac{p_a^2}{a^{2n-4}}\right>_L=\text{Re}\left[\frac{\Psi^{\ast}\left(\hat{p}_a^2/a^{2n-4}\right)_W\Psi}{\Psi^{\ast}\Psi}\right]=-\frac{1}{a^{2n-4}\Theta}\frac{\partial^2\Theta}{\partial a^2}+\frac{1}{a^{2n-4}}\left(\frac{\partial S}{\partial a}\right)^2+\frac{2n-4}{a^{2n-3}\Theta}\frac{\partial \Theta}{\partial a}-\frac{(2n-4)(2n-3)}{4a^{2n-2}},
\label{lev}
\end{equation}
where $\Psi=\Theta e^{iS}$.

The time rate of the momentum $\dot{p}_a$ can be found through the equation \cite{holland}
\begin{equation}
\dot{p}_a=-\frac{\partial Q}{\partial a},
\end{equation}
where $Q$ is the quantum mechanical potential \cite{bohm} [see Eq. (\ref{eq. universe})]
\begin{equation}
Q=-\frac{1}{4(n-1)(n-2)a^{n-4}}\frac{1}{\Theta}\frac{\partial^2\Theta}{\partial a^2}.
\end{equation}

Substituting Eq. (\ref{polar form}) with $\alpha=1/(n-1)$ into the above equation we have
\begin{equation}
Q=\frac{n-1}{n-2}\frac{\gamma}{\gamma^2+t^2}-\frac{4\gamma^2}{(\gamma^2+t^2)^2}\frac{n-1}{n-2}a^{n-2}
\end{equation}
which implies 
\begin{equation}
\dot{p}_a=\frac{4\gamma^2(n-1)}{(\gamma^2+t^2)^2}a^{n-3}.
\end{equation}

Substituting Eq. (\ref{polar form}) with $\alpha=1/(n-1)$ into Eq. (\ref{lev}) we have
\begin{equation}\begin{aligned}
\left<\frac{p_a^2}{a^{2n-4}}\right>_L=&-\frac{1}{a^{2n-3}}\left\{-\frac{4\gamma(n-1)^2}{\gamma^2+t^2}a^{n-3}+\left[\frac{4\gamma(n-1)}{\gamma^2+t^2}\right]^2a^{2n-5}\right\}+\frac{1}{a^{2n-4}}\left[\frac{4t(n-1)}{\gamma^2+t^2}a^{n-3}\right]^2\\&+\frac{2n-4}{a^{2n-2}}\left[1-\frac{4\gamma(n-1)}{\gamma^2+t^2}a^{n-2}\right]-\frac{(2n-4)(2n-3)}{4a^{2n-2}}.
\end{aligned}\end{equation}

Therefore
\begin{equation}
\begin{aligned}
&\left<R^{\hat{i}}_{\phantom{\hat{i}}\hat{0}\hat{i}\hat{0}}\right>_L=-\frac{2\gamma^2}{(\gamma^2+t^2)^2}\frac{1}{n-2}\frac{1}{a^2}+\frac{n-3}{4(n-1)^2(n-2)^2}\left<\frac{p_a^2}{a^{2n-4}}\right>_L\\
&\left<R^{\hat{i}}_{\phantom{\hat{i}}\hat{j}\hat{i}\hat{j}}\right>_L=\frac{1}{4(n-1)^2(n-2)^2}\left<\frac{p_a^2}{a^{2n-4}}\right>_L.
\end{aligned}
\end{equation}

The Ricci scalar is given in terms of the components of the curvature tensor in the tetrad basis by
\begin{equation}
\left<R\right>_L=-2(n-1)\left<R^{\hat{i}}_{\phantom{\hat{i}}\hat{0}\hat{i}\hat{0}}\right>_L+(n-1)(n-2)\left<R^{\hat{i}}_{\phantom{\hat{i}}\hat{j}\hat{i}\hat{j}}\right>_L.
\end{equation}

\begin{figure}[!htb]
\centering
\includegraphics[scale=0.8]{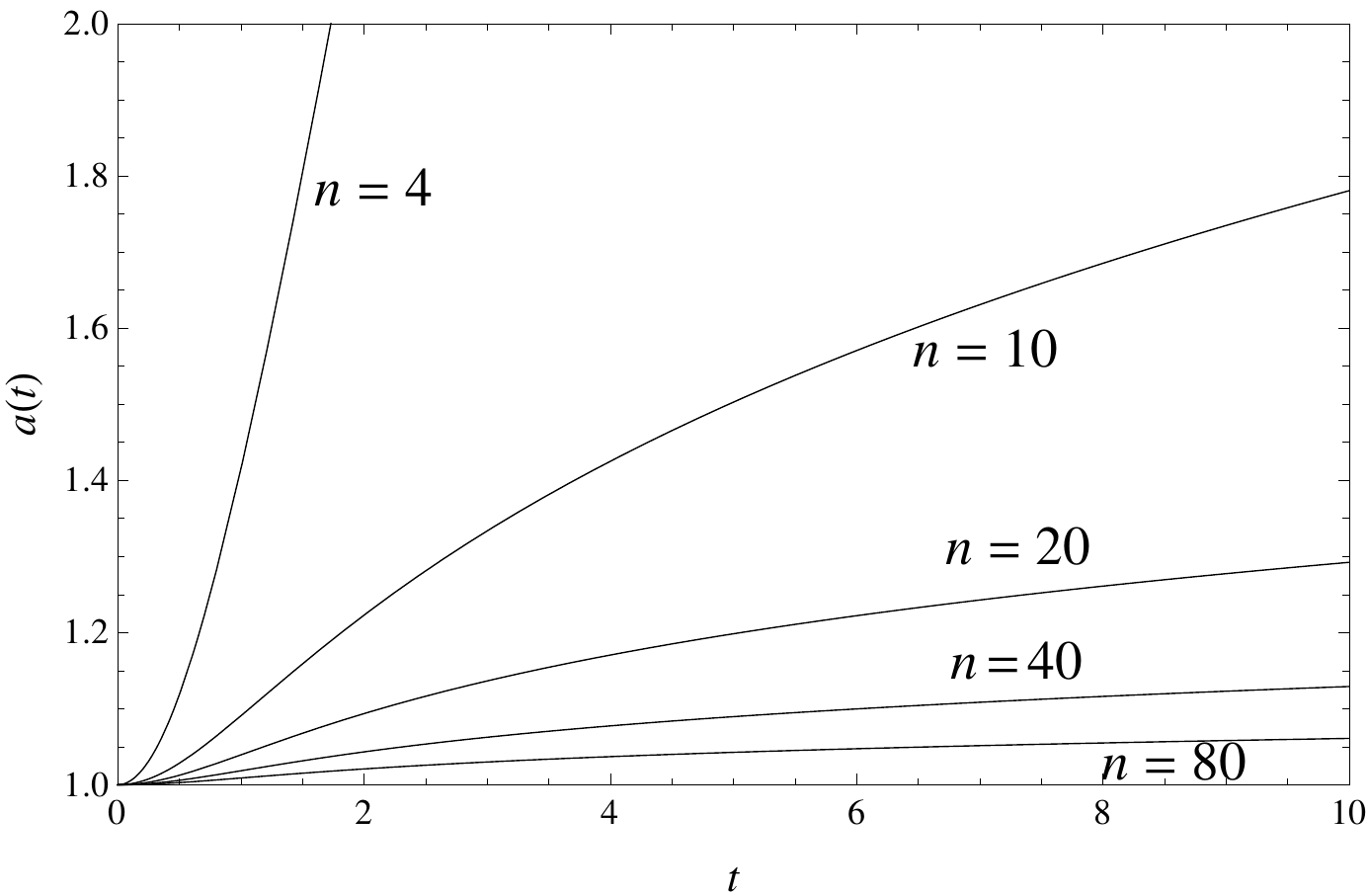}
\caption{The Bohmian trajectory of the scale factor for $n=4,10,20,40$ and $80$, for $a_0=\gamma=1$. The expansion of the universe slower  as $n$ grows up and as $n\to \infty$ we approach a static universe.}
\label{fig a(t)}
\end{figure}

Let us now present some graphics relating the dimension of the FLRW universe. We start with the graphic of the scale factor in the de Broglie-Bohm interpretation of quantum mechanics, Fig. \ref{fig a(t)} (remember that the scale factor in the many-worlds and in the de Broglie-Bohm interpretation differs only by an overall constant factor). For $\alpha=\frac{1}{n-1}$ the scale factor behaves as [see Eq. (\ref{a(t)})]
\begin{equation}
a(t)=a_0(\gamma^2+t^2)^{\frac{1}{n-2}}
\end{equation} 
in the conformal gauge. Note that for $n\to\infty$ the scale factor approaches the constant $a(t)=a_0$ and that the expansion is slower as $n$
grows up.

\begin{figure}[!htb]
\centering
\includegraphics[scale=0.8]{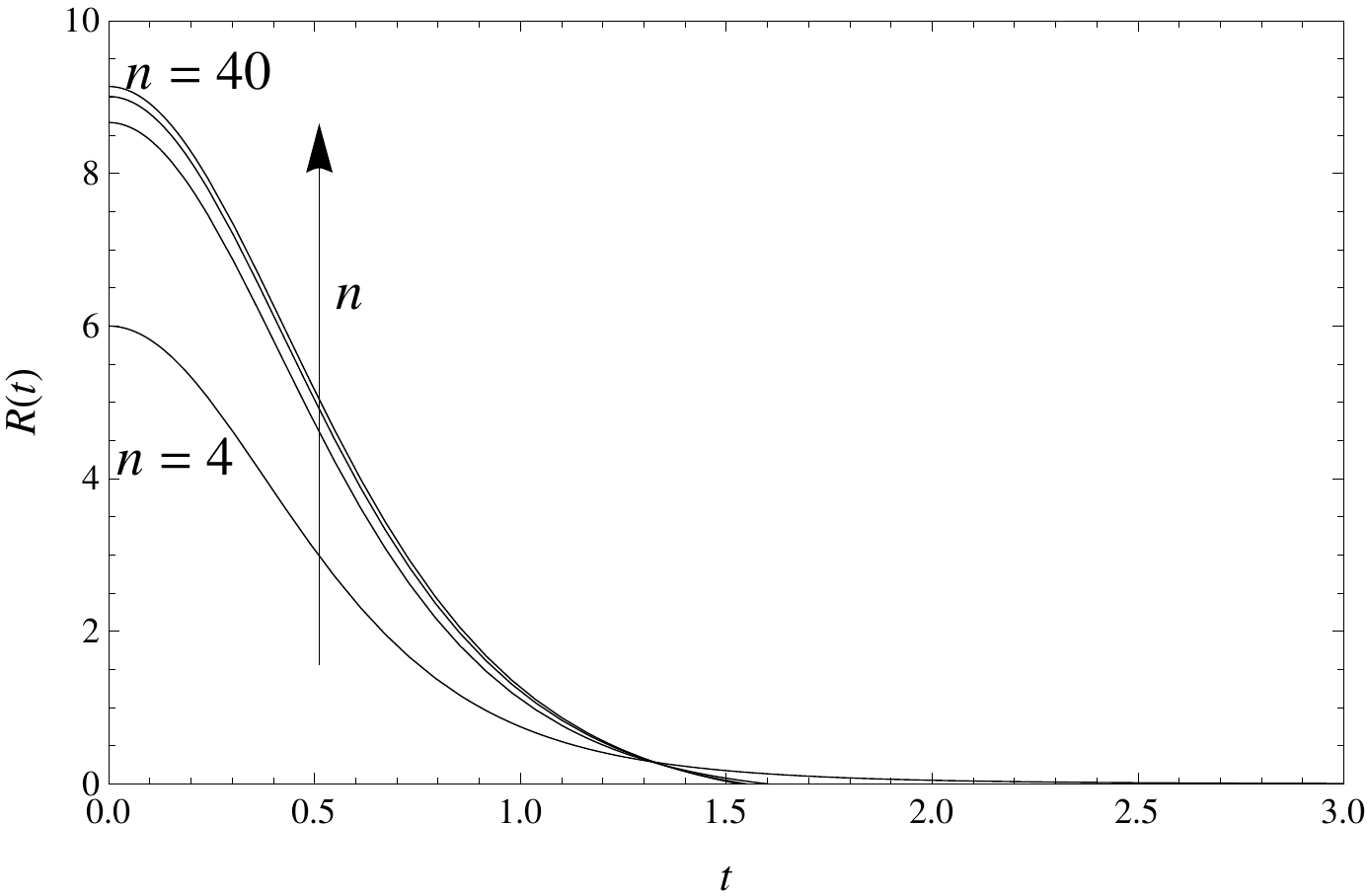}
\caption{The local expectation value of the Ricci scalar $R(t)$ for $n=4,10,20,40$ ($a_0=\gamma=1$). They are perfectly regular for all values of $t$ indicating that the singularity is excluded by the introduction of quantum cosmology.}
\label{R(t)}
\end{figure}

The graphic for the local expectation value of the Ricci scalar,  which is perfectly regular for all $n$, is shown in Fig. \ref{R(t)}.

With the components of the Riemann curvature tensor in the tetrad basis we can find the energy density and pressure of the fluid filling the universe through the Einstein equations
\begin{equation}
T^{\alpha}_{\phantom{\alpha}\beta}=\frac{1}{k_n}\left(R^{\alpha}_{\phantom{\alpha}\beta}-\frac{1}{2}\delta^{\alpha}_{\phantom{\alpha}\beta}R\right)=\text{diag}(-\rho,p,\dots,p).
\end{equation}

\begin{figure}[!htb]
\centering
\includegraphics[scale=0.8]{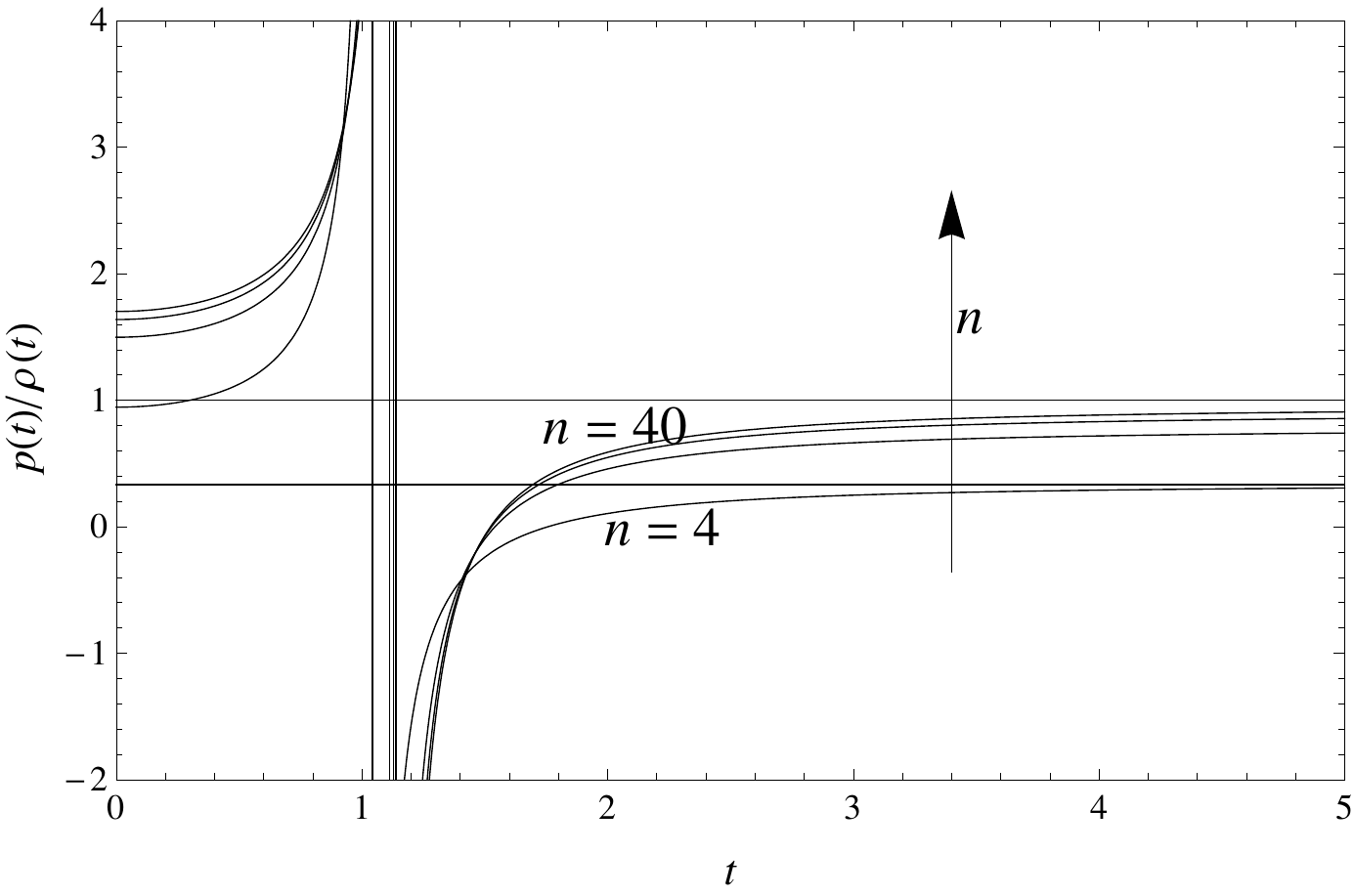}
\caption{The ratio $\left<p\right>_L/\left<\rho\right>_L$ for  $n=4,10,20$,$40$  and $a_0=\gamma=1$. In four dimensions the classical behavior of the fluid is recovered while $n\to \infty$ $\left<p\right>_L/\left<\rho\right>_L\to 1$.}
\label{ratio}
\end{figure}

Therefore
\begin{equation}
\begin{aligned}
&\rho=\frac{1}{k_n}\left[\frac{(n-1)(n-2)}{2}R^{\hat{i}}_{\phantom{\hat{i}}\hat{j}\hat{i}\hat{j}}\right]\\
&p=\frac{1}{k_n}\left[(n-2)R^{\hat{i}}_{\phantom{\hat{i}}\hat{0}\hat{i}\hat{0}}-\frac{(n-2)(n-3)}{2}R^{\hat{i}}_{\phantom{\hat{i}}\hat{j}\hat{i}\hat{j}}\right],
\end{aligned}\end{equation}
and we can plot the ratio $\left<p\right>/\left<\rho\right>$ as shown in Fig. \ref{ratio}. Wee see that in the classical limit, where quantum effects become negligible, $\left<p\right>/\left<\rho\right>\to1/3$ in four-dimensions as expected. For other dimensions $\left<p\right>/\left<\rho\right>$ does not tend to the classical limit and we can also see that as $n\to\infty$, $\left<p\right>/\left<\rho\right>\to1$. In this sense four-dimensions are privileged among the others. The introduction of quantum cosmology in four-dimensions reproduces the classical behavior of the fluid.

\begin{figure}[!htb]
\centering
\includegraphics[scale=0.8]{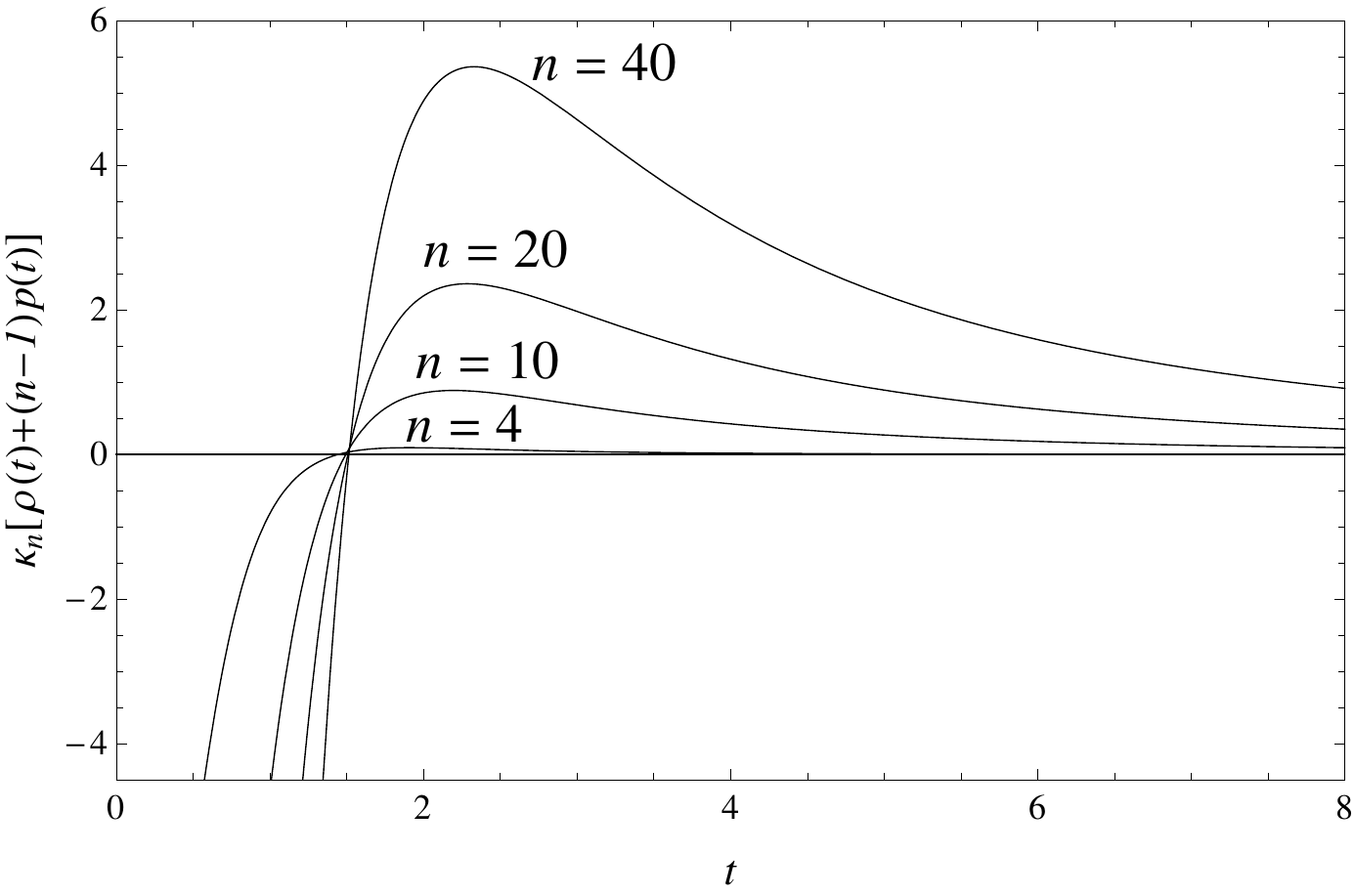}
\caption{The graphic of $\kappa_n[\left<\rho\right>_L+(n-1)\left<p\right>_L]$ for $n=4,10,20$ and $40$ . We see that the strong energy condition is violated when $t$ is small and returns to which is classically expected as $t$ gets bigger.}
\label{energycondition}
\end{figure}

We can also study the behavior of the quantity $\kappa_n[\left<\rho\right>_L+(n-1)\left<p\right>_L]$, which comes from the strong energy condition. We see in Fig. \ref{energycondition} that the strong energy condition, which demands that $\rho+(n-1)p\geq 0$, is violated for small $t$. This is expected since the singularities have been removed. But note that for large $t$ the quantity $\rho+(n-1)p$ is greater than zero, returning to the classical behavior.

\section{conclusions}

The time dependence of the scale factor in and n-dimensional FLRW universe with a perfect fluid was found by two different ways. By solving Einstein equations and via Hamiltonian formalism, where we extended Schutz formalism to dimensions other than four. By comparing both results we could test the validity of this extrapolation. We then found that the big-bang singularity is still present in the n-dimensional FLRW universe. By solving the Wheeler-deWitt equation for the universe, we obtained by the superposition principle a finite norm wave packet for the universe and we could find the expectation value, in the many-worlds interpretation, and the trajectory in  de Broglie-Bohm interpretation, of the scale factor and discovered that the big-bang singularity is removed with the introduction of quantum cosmology.

By quantizing the independent components of the curvature tensor in the tetrad basis, we showed graphically that the local expectation value of the Ricci scalar is well behaved for all time, giving one more indication that the big-bang singularity has been removed. We also studied the quantum behavior of the ratio $p/\rho$ and showed that this ratio tends to the classical limit when $t\to\infty$ only in four-dimensions. This indicates that, in a certain way, four dimensions are single out  from the view point of  the quantum cosmology since we recover classical quantities as the quantum behavior becomes negligible. We also showed that for any dimension the strong energy condition is violated as expected since we do not have singularities.

\acknowledgements 
J.P.M. Pitelli thanks FAPESP for financial support. P.S. Letelier aknowledges FAPESP and CNPq for partial financial support.

%-------------------------------------------------------------------------------The End-----------------------------------------------------------------

\end{document}